\documentclass[a4paper]{article}

\usepackage{INTERSPEECH2021}

\usepackage{multirow}
\usepackage{url}
\usepackage{cite}

\title{ECAPA-TDNN Embeddings for Speaker Diarization}
\name{Nauman Dawalatabad$^{1,2}$\thanks{This work is done when N. Dawalatabad was at Mila - Quebec Artificial Intelligence Institute, Canada, and H. Na was at SAIT AI Lab, Montreal.}, Mirco Ravanelli$^2$, Fran\c{c}ois Grondin$^3$, \\ Jenthe Thienpondt$^4$, Brecht Desplanques$^4$, Hwidong Na$^5$}

\address{
  $^1$Indian Institute of Technology Madras, India\\
  $^2$Mila - Quebec Artificial Intelligence Institute, Canada\\
  $^3$Universit\'e de Sherbrooke, Canada\\
  $^4$IDLab,  Ghent University - imec, Belgium\\
  $^5$Samsung Advanced Institute of Technology, Suwon, South Korea }

\email{nauman@cse.iitm.ac.in, mirco.ravanelli@gmail.com}

\begin{document}

\maketitle
\begin{abstract}
Learning robust speaker embeddings is a crucial step in speaker diarization. Deep neural networks can accurately capture speaker discriminative characteristics and popular deep embeddings such as x-vectors are nowadays a fundamental component of modern diarization systems. Recently, some improvements over the standard TDNN architecture used for x-vectors have been proposed. The ECAPA-TDNN model, for instance, has shown impressive performance in the speaker verification domain, thanks to a carefully designed neural model. 

In this work, we extend, for the first time, the use of the ECAPA-TDNN model to speaker diarization. Moreover, we improved its robustness with a powerful augmentation scheme that concatenates several contaminated versions of the same signal within the same training batch. The ECAPA-TDNN model turned out to provide robust speaker embeddings under both close-talking and distant-talking conditions. Our results on the popular AMI meeting corpus show that our system significantly outperforms recently proposed approaches. 
  
\end{abstract}

\section{Introduction}

Speaker diarization answers the question of ``\textit{who spoke when?}" in a given conversation~\cite{xavier12-review,park2021review}.
Diarization is used in many conversational AI systems and applied in various domains such as telephone conversations, broadcast news, meetings, clinical recordings, and many more~\cite{park2021review}.
Modern diarization systems rely on neural speaker embeddings coupled with a clustering algorithm.

Despite the recent progress, speaker diarization is still one of the most challenging speech processing tasks~\cite{sell-dihard2018}. 
Research in this field is very active, and it is fostered by popular challenges such as DIHARD~\cite{sell-dihard2018}.  
As for clustering, various approaches have been proposed in the literature, including top-down and bottom-up agglomerative clustering~\cite{xavier12-review}. Spectral clustering, which is a graph clustering method based on the eigenanalysis of the Laplacian matrix, has recently shown promising performance on speaker diarization \cite{park2021review,park_autotune,pal21-meta}. 

Several research efforts have been devoted to neural speaker embeddings as well. Modern speaker embeddings such as d-vectors~\cite{dvector-1}, c-vectors~\cite{cvector}, and x-vectors~\cite{xvector} have shown to capture speaker discriminative characteristics very well. The x-vector model, for instance, is based on a Time Delay Neural Network (TDNN) and is now a fundamental component of the state-of-the-art diarization systems~\cite{park2021review}.

In this paper, we propose to use an enhanced version of the standard TDNN model based on Emphasized Channel Attention, Propagation, and Aggregation (ECAPA-TDNN)~\cite{ecapa-tdnn}.
The ECAPA-TDNN model employs a channel- and context-dependent attention mechanism, Multilayer Feature Aggregation (MFA), as well as Squeeze-Excitation (SE) and residual blocks.
This model has recently shown impressive performance in the speaker verification domain~\cite{ecapa-tdnn}.
It has shown the best performance in the text-independent task of SdSV Challenge on short-duration speaker verification~\cite{sdsv20,Thienpondt2020}. 
This makes it a good choice for speaker diarization, as speaker turns in realistic conversations can be of short duration.
This is the first time that the ECAPA-TDNN model is used in the context of speaker diarization.
Our work further improves its robustness by training the ECAPA-TDNN model with an extensive on-the-fly augmentation scheme that relies on the combination of different techniques for speech contamination.
We propose to use augmentation techniques such as waveform dropout, frequency dropout, speed perturbation, reverberation, and additive noise while training the ECAPA-TDNN model.
Differently from standard augmentation schemes, all the contaminated versions are concatenated within the same training batch.

We conducted our experimental studies using the popular AMI~\cite{ami} meeting dataset considering different types of audio streams. 
The proposed system shows highly competitive performance and overtakes recent approaches in speaker diarization.
To foster replicability, we made the code and the pre-trained models available in the SpeechBrain project\footnote{\url{https://speechbrain.github.io/}}.

\section{ECAPA-TDNN Diarization}
\label{sec:proposed}
In this section, we describe the various modules involved in the proposed ECAPA-TDNN based speaker diarization system.

\begin{figure}
    \centering
    \includegraphics[scale=0.27]{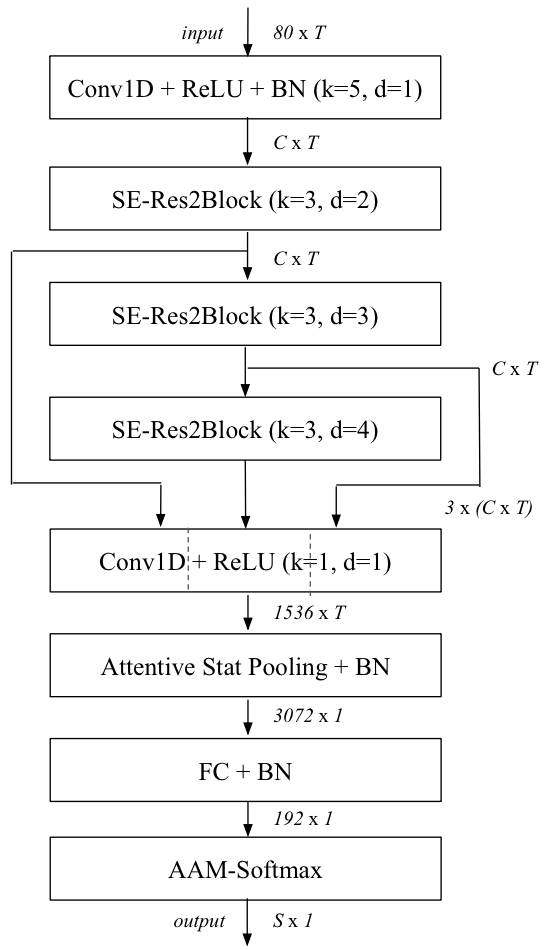}
    \caption{Block diagram of the ECAPA-TDNN model~\cite{ecapa-tdnn}.}
    \label{fig:block}
\end{figure}

\subsection{Speaker embeddings}
\label{subsec:ecapa}

Modern speaker embeddings are computed from neural models trained to classify speaker identities from a large pool of speakers~\cite{xvector, ecapa-tdnn, magneto}.
A temporal statistics pooling layer is used to map the variable length input to a fixed-length representation.
After training, the fixed-length speaker embeddings are extracted from the activations of the penultimate layer in the network. 

As shown in Figure~\ref{fig:block}, the ECAPA-TDNN~\cite{ecapa-tdnn} model architecture is based on the popular x-vector topology \cite{xvector} and it introduces several enhancements to create more robust speaker embeddings. The pooling layer uses a channel- and context-dependent attention mechanism, which allows the network to attend different frames per channel. 
1-dimensional Squeeze-Excitation (SE)~\cite{se_block} blocks rescale the channels of the intermediate frame-level feature maps to insert global context information in the locally operating convolutional blocks. Next, the integration of 1-dimensional Res2-blocks~\cite{res2net} improves performance while simultaneously reducing the total parameter count by using grouped convolutions in a hierarchical way. Finally, Multi-layer Feature Aggregation (MFA)~\cite{multi_stage_aggregation} merges complementary information before the statistics pooling by concatenating the final frame-level feature map with intermediate feature maps of preceding layers. 

The network is trained by optimizing the AAM-softmax~\cite{arcface} loss on the speaker identities in the training corpus. The AAM-softmax is a powerful enhancement compared to the regular softmax loss in the context of fine-grained classification and verification problems. It directly optimizes the cosine distance between the speaker embeddings. As a consequence, complex scoring backends such as Probabilistic Linear Discriminant Analysis (PLDA)~\cite{plda} can be avoided.

\subsection{Data augmentation}
\label{subsec:aug}
Data augmentation is a common approach to improve the robustness of a neural model. Speech can be contaminated in different ways. In this study, we train the ECAPA-TDNN model with the following augmentation strategies: 

\begin{itemize}
    \item \textit{Waveform dropout:} replaces some random chunks of the original waveform with zeros~\cite{specaugment}.
    \item \textit{Frequency dropout:} filters the original signal with random band-stop filters to add zeros in the frequency spectrum. \cite{specaugment}.
    \item \textit{Speed Perturbation:} resamples the audio signal to a sampling rate that is slightly different from the original one. With this simple trick, we can synthesize a speech signal that sounds a bit faster or slower than the original one. 
    This is useful as the speaking rate may vary within and across speakers~\cite{dawalatabad20-novel}.
    To avoid changing the speaker characteristics significantly, we restrict the speed perturbation to a maximum of $\pm 5\%$. 
    \item \textit{Reverberation:} introduces reverberation by convolving the signal with a randomly selected room impulse response.
    \item \textit{Additive Noise:}  adds a randomly selected noise sequence to the speech signal with a random signal-to-noise ratio.
    \item \textit{Noise + Reverberation:} combines the noise and reverberation disturbances.
\end{itemize}

All of these augmentations are applied on-the-fly to every speech sentence processed by the neural network. This way, we generate a different contaminated data at every epoch. 
In standard augmentation pipelines, the signal is contaminated with one or more augmentation strategies and then used for training a neural network.  Instead, in this work, we propose to concatenate the original speech signal with all the contaminated versions produced by the aforementioned contamination techniques~\cite{hoffer2019augment}.
This way, within each training batch, our model observes the same sentence corrupted in six different ways. These different ``views'' of the same signal force the gradient to point to a direction of the parameter space that is inherently robust against signal variations, thus proving an important regularization effect.

\subsection{Spectral clustering}
\label{sec:spec_clust}
Spectral Clustering (SC) is a popular  clustering approach for speaker diarization that has recently shown highly competitive performance compared to the traditional Agglomerative Hierarchical Clustering (AHC) with PLDA backend~\cite{park_autotune,pal21-meta}.

There are multiple methods to perform spectral clustering~\cite{spec_tutorial}.
We follow the unnormalized spectral clustering approach similar to~\cite{park2021review,spec_tutorial,park_autotune}.
The affinity matrix $A$ is calculated using the cosine similarity metric.
It is important to prune out the smaller values in $A$ to focus more on prominent values in the matrix.
Similar to~\cite{spec_tutorial}, we use the actual similarity values in affinity matrix while calculating the Laplacian matrix.
An unnormalized Laplacian matrix is estimated using the symmetrized $A$ as done in~\cite{park2021review,spec_tutorial}.
The Laplacian matrix is subjected to eigendecomposition. 
We estimate the number of speakers $k$ using the maximum eigengap approach~\cite{spec_tutorial}. Next, we compute the first $k$ eigenvectors. The rows of the eigenvector matrix are the $k$ dimensional spectral embeddings corresponding to each analyzed speech segment. The estimated spectral embeddings that are expected to be more separable than the original speaker embeddings, are clustered using the standard $k$-means algorithm.

\section{Experimental Setup}
\label{sec:exp}

\subsection{Datasets}
The ECAPA-TDNN model is trained with VoxCeleb1 and VoxCeleb2 data~\cite{vox1,vox2}.
The RIRs\footnote{https://www.openslr.org/28/} and MUSAN~\cite{musan2015} datasets  are used for data-augmentation purposes. 

For diarization,  we use the Augmented Multi-party Interaction (AMI)  meeting dataset~\cite{ami}. We use the official ``Full ASR corpus" split with TNO meetings excluded from the Dev and Eval set. Official manual annotations serve as the ground truth for evaluation. 
The channels in the microphone array are beamformed with the standard BeamformIt toolkit~\cite{xavier07-beamformit}.
The same split is used by many other works~\cite{cvector,pal21-meta,landini2020VBX,li2020discriminative,pal2019speaker}.

\subsection{Speaker embeddings}
The ECAPA-TDNN model is fed with 80-dimensional log Mel filterbank energies that are mean normalized per input segment. The model parameter updates are determined by the Adam~\cite{adam} optimizer with a Cyclical Learning Rate (CLR)~\cite{clr} using a triangular policy. Training is done for 10 epochs with batches of 32 segments. The original batch is augmented in six different ways, leading to an equivalent batch size of 192. The contamination with MUSAN additive noise is done with a random Signal-to-Noise Ratio (SNR)  ranging from 0 to 10 dB. Reverberation is added by convolving with a random impulse response from the aforementioned RIRs dataset.

We train the model with 3~sec random crops of the speaker utterances. The architectural hyper-parameters of the ECAPA-TDNN model are the same as in~\cite{ecapa-tdnn}.
The model achieves a promising EER of 0.69\% and a minDCF of 0.0826 on the original VoxCeleb1 (cleaned) verification set. 
Additional details can be found in the Voxceleb recipe in SpeechBrain~\cite{SB2021}.

\subsection{Diarization setup}
The embeddings of each continuous speech segment are extracted with a sliding window of size 3~sec and a shift of 1.5~sec.
The maximum number of estimated speakers is set to 10. The pruning threshold for the affinity matrix is determined on the AMI Dev set.

We use the standard Diarization Error Rate (DER) as evaluation metric~\cite{DER-xavier-blog}. 
The DER consists of a Speaker Error Rate (SER), False Alarm (FA), and Missed Speech (MS) component.
The SER represents the errors introduced due to incorrect labeling of speaker segments.
The FA and MS occur due to errors introduced by the Voice Activity Detection (VAD) system. 
Since this work focuses on improving the clustering module, similar to~\cite{pal21-meta} we use oracle speech/non-speech labels from the ground truth.
To enable direct comparison with \cite{pal21-meta}, we do not use any realignment post-processing step. 
Similar to \cite{pal21-meta}, a forgiveness collar of 0.25~sec is used and the speaker overlap regions are ignored during scoring (as also standard by NIST). We use the standard NIST evaluation tool available in SpeechBrain~\cite{SB2021}.

\section{Results}

\subsection{Baseline systems}

We compare our system with the recently proposed ClusterGAN~\cite{pal21-meta}, MCGAN~\cite{pal21-meta} and Variational Bayes~\cite{landini2020VBX} based diarization techniques.

The approaches mentioned in~\cite{pal21-meta} are sophisticated systems where ClusterGAN and MCGAN models are trained using x-vectors extracted from TDNN~\cite{xvector}.
The embeddings obtained from the models are fused with x-vectors and then clustered using a spectral clustering algorithm.
The number of speakers is estimated with the Normalized Maximum Eigengap (NME-SC) technique~\cite{park_autotune}. %
The NME-SC algorithm is expensive in terms of runtime because it requires iterating through multiple values of the pruning threshold to find the best setting. 
Therefore, this work estimates the number of speakers with the standard maximum eigengap criterion proposed in \cite{spec_tutorial}.
We also compare our system with the Variational Bayes (VBx) approach~\cite{landini2020VBX} under the same experimental setup as used in this paper. VBx uses ResNet101-based x-vector embeddings  \cite{resnet,BUT_resnet_xvector} that are clustered using Bayesian Hidden Markov Model (BHMM) \cite{bhmm, BUT_resnet_xvector,nagrani2020voxsrc}.

All three approaches have shown competitive or state-of-the-art performance (to the best of our knowledge) on the AMI dataset, making them strong baselines for comparison.

\begin{table}[]
\centering
\caption{Diarization Error Rates (DERs) on AMI dataset using the beamformed array signal on baseline and proposed systems -- less is better.}

\resizebox{0.48\textwidth}{!}{%
\begin{tabular}{@{} c c c c c c @{}}
\toprule
\multirow{2}{*}{Embedding}  & \multirow{2}{*}{Back-end} & \multicolumn{2}{c}{\begin{tabular}[c]{@{}c@{}}Oracle \\ num of speakers\end{tabular}} & \multicolumn{2}{c}{\begin{tabular}[c]{@{}c@{}}Estimated \\ num of speakers\end{tabular}} \\ \cmidrule(l){3-6} 
                            &                           & Dev                                        & Eval                                      & Dev                                         & Eval                                        \\ \midrule
xvector+ClusterGAN          & \multirow{2}{*}{k-means \cite{pal21-meta}}  & 6.62                                       & 6.46                                      & 9.57                                        & 8.63                                        \\ \cmidrule(r){1-1} \cmidrule(l){3-6} 
xvector+MCGAN               &                           & 5.64                                       & 5.48                                      & 6.47                                        & 8.76                                        \\ \midrule
xvector+ClusterGAN          & \multirow{2}{*}{SC \cite{pal21-meta}}       & 3.93                                       & 3.60                                      & 6.21                                        & 2.87                                        \\ \cmidrule(r){1-1} \cmidrule(l){3-6} 
xvector+MCGAN               &                           & 5.49                                       & 4.23                                      & 5.02                                        & 4.92                                        \\ \midrule
xvector (ResNet101)         & VBx \cite{landini2020VBX}                       & -                                          & -                                         & 4.27                                        & 4.58                                        \\ \midrule
\multicolumn{6}{c}{Proposed Approach}                                                                                                                                                                                                      \\ \midrule
\multirow{2}{*}{ECAPA-TDNN} & k-means                   & 3.03                                       & 3.69                                      & 4.65                                        & 5.10                                        \\ \cmidrule(l){2-6} 
                            & SC                        & \textbf{2.82}                                       & \textbf{2.65}                                      & \textbf{3.66}                                        & \textbf{3.01}                                        \\ \bottomrule
\end{tabular}
}
\label{tab:der-main}
\end{table}

\subsection{Comparison with baseline systems}
Table~\ref{tab:der-main} compares the aforementioned baselines with the proposed ECAPA-TDNN diarization system. The DERs reported in the table are estimated on the beamformed audio in two scenarios, i.e., (i) when the number of speakers is known before diarization (oracle number of speakers), and (ii) when the number of speakers is not known apriori and has to be automatically estimated.
We estimate the number of speakers with maximum eigengap criterion for both, k-means and Spectral Clustering~(SC) backends.

From Table~\ref{tab:der-main} it emerges that the proposed approach significantly outperforms the baselines in most of the cases.
There is a significant improvement compared to the baseline systems with SC as a backend. 
The only exception is the performance of  x-vector+ClusterGAN embeddings with an unknown number of speakers, for which the DER of 2.87\% on the Eval set is comparable to the DER of 3.01\% achieved by our system. In all the other cases, the proposed system significantly outperforms the x-vector+ClusterGAN based system.
Our system also outperforms x-vector+MCGAN  based systems in all the cases.
With an SC backend, relative improvements of 37.4\% (from 4.23\% to 2.65\%) and 38.8\% (from 4.92\% to 3.01\%) are observed on the Eval set for oracle and estimated number of speakers, respectively.
This improvement is 34.3\% (from 4.58\% to 3.01\%) with respect to the VBx system for an unknown number of speakers.

Interestingly, the ECAPA-TDNN diarization system achieves a noteworthy performance also with a simple k-mean clustering backend. 
For instance, comparing with the x-vector+ClusterGAN system, our system shows improvement of 42.9\% (from 6.46\% to 3.69\%) and 40.9\% (from 8.63\% to 5.10\%) on Eval set for oracle and estimated number of speakers, respectively.
 Compared to the x-vector+MCGAN based system, our system show improvements of 32.7\% (from 5.48\% to 3.69\%) and 41.8\% (from 8.76\% to 5.10\%) on the Eval set for oracle and estimated number of speakers, respectively.
This further confirms the robustness of the proposed embeddings.

\begin{table}[]
\centering
\caption{DERs comparison between x-vectors and ECAPA-TDNN with SC backend on beamformed data -- less is better.}
\resizebox{0.42\textwidth}{!}{%
\begin{tabular}{@{}c c c c c @{}}
\toprule
\multirow{2}{*}{Embedding} & \multicolumn{2}{c }{\begin{tabular}[c]{@{}c@{}}Oracle \\ num of speakers\end{tabular}} & \multicolumn{2}{c }{\begin{tabular}[c]{@{}c@{}}Estimated\\ num of speakers\end{tabular}} \\ \cmidrule(l){2-5} 
                       & Dev                                        & Eval                                      & Dev                                        & Eval                                        \\ \midrule
x-vector              & 6.14                                       & 8.57                                      & 9.21                                       & 11.04                                       \\ 
ECAPA-TDNN              & \textbf{2.82}                                       & \textbf{2.65}                                      & \textbf{3.66}                                       & \textbf{3.01}                                       \\ \bottomrule
\end{tabular}
}
\label{tab:xvect-der}
\end{table}

\subsection{Comparison with x-vector embedding}
Table~\ref{tab:xvect-der} shows the DERs obtained using standard x-vector embeddings on beamformed data with SC as backend. The x-vector embeddings used here are trained with the same augmentation scheme described in Section \ref{subsec:aug}.

The diarization performance with ECAPA-TDNN embeddings is far superior to that achieved by standard x-vector embeddings.
This is due to various improvements introduced by the ECAPA-TDNN model. 
The same trend is observed on the other audio streams (individual distant mics, HeadsetMix, LapelMix).

\subsection{Ablation study on augmentation}
To confirm the effectiveness of the proposed augmentation approach, we conducted experiments by training the ECAPA-TDNN model with no data augmentation and with standard data augmentation techniques. The standard augmentation is performed by applying the aforementioned contamination methods to each training sentence (including a clean version of the signal). 
The proposed augmentation, instead, uses the batch construction technique described in Section~\ref{subsec:aug}.

It can be seen from Table~\ref{tab:aug-der}  that the performance achieved with the proposed augmentation technique is better than that of a standard augmentation.
The same trend is observed on all other audio streams.
The DER is even worse when no data augmentation is used. 
The special batch construction technique adopted in the proposed augmentation scheme clearly helps to improve the robustness of the diarization system.  


\begin{table}[t!]
\centering
\caption{DERs achieved when comparing different augmentation techniques with SC backend on beamformed data -- less is better.}
\resizebox{0.42\textwidth}{!}{%
\begin{tabular}{@{}ccccc@{}}
\toprule
\multirow{2}{*}{Augmentation} & \multicolumn{2}{c}{\begin{tabular}[c]{@{}c@{}}Oracle \\ num of speakers\end{tabular}} & \multicolumn{2}{c}{\begin{tabular}[c]{@{}c@{}}Estimated \\ num of speakers\end{tabular}} \\ \cmidrule(l){2-5} 
                              & Dev                                        & Eval                                      & Dev                                         & Eval                                        \\ \midrule
Without  Aug.                 & 3.95                                       & 3.92                                      & 5.72                                        & 7.45                                        \\ 
Standard Aug.                 & 3.04                                       & 2.64                                      & 4.48                                        & 4.51                                        \\ 
Proposed Aug.                 & \textbf{2.82}                                       & \textbf{2.65}                                      & \textbf{3.66}                                        & \textbf{3.01}                                        \\ \bottomrule
\end{tabular}
}
\label{tab:aug-der}
\end{table}

\subsection{Distant and close talking  microphones}
Table~\ref{tab:mics-der} reports the performance of the proposed system achieved with the different microphone settings, including distant microphones, HeadsetMix, and LapelMix audio streams. 
The DERs obtained on Headsetmix and LapelMix audio streams are reported to facilitate the comparison with other works.
The signals in the HeadsetMix and LapelMix streams are relatively clean, and hence the results on these streams give an estimate of the lowest DERs that can be achieved by the proposed system. 
In the case of the oracle number of speakers, the best DERs achieved on the HeadsetMix signal are 2.02\%  and 1.78\% for Dev set and Eval set, respectively.
For the LapelMix audio, the best DERs of 2.34\% and 2.57\% on the Dev and the Eval set are observed for an unknown number of speakers.

The row \textit{Distant-Mic} in Table~\ref{tab:mics-der} reports the average DER over the eight distant microphones of the AMI microphone array. 
It is interesting to note that the average DERs on the distant mics is comparable to the performance obtained on the beamformed audio.
It is also worth noticing that the performance degradation observed when switching from distant-talking conditions (e.g., beamformed or distant microphone) to close-talking ones (e.g., HeadsetMix and LapelMix) is not huge. Similar behavior was also observed with the k-means backend.
These results can be attributed to the data augmentation scheme. With the proposed augmentation approach, we indeed train the neural network with different environmental contamination effects (noise, reverberation, and noise + reverberation) and we thus implicitly achieve robustness in different acoustic conditions.

\begin{table}[]
\centering
\caption{DERs achieved by the proposed system on distant and close talking audio streams with SC backend -- less is better.}
\begin{tabular}{@{} c c c c c @{}}
\toprule
\multirow{2}{*}{Audio Streams}   & \multicolumn{2}{c}{\begin{tabular}[c]{@{}c@{}}Oracle \\ num of speakers\end{tabular}} & \multicolumn{2}{c}{\begin{tabular}[c]{@{}c@{}}Estimated \\ num of speakers\end{tabular}} \\ \cmidrule(l){2-5} 
                                 & Dev                                        & Eval                                      & Dev                                         & Eval                                        \\ \midrule
HeadsetMix                       & 2.02                                       & 1.78                                      & 2.43                                        & 4.03                                        \\ 
LapelMix                         & 2.17                                       & 2.36                                      & 2.34                                        & 2.57                                        \\ 
Distant-Mic (avg.)               & 2.81                                       & 3.12                                      & 3.33                                        & 3.75                                        \\ 
Beamformed & 2.82                                       & 2.65                                      & 3.66                                        & 3.01                                        \\ \bottomrule
\end{tabular}
\label{tab:mics-der}
\end{table}

\section{Conclusion and Future Work}
\label{sec:conc}
In this paper, we improved speaker diarization performance on AMI meeting data by focusing on two critical components of the speaker diarization pipeline. (i) A better data augmentation technique which includes waveform dropout, frequency dropout, speed perturbation, reverberation, and adding noise disturbances. Crucially, these multiple augmented views of the signal are gathered in the same batch. (ii) The use of the ECAPA-TDNN model for the extraction of more robust speaker embeddings.
The proposed approach has surpassed the performance of the most recent techniques such as ClusterGAN, MCGAN, and VBx. Moreover, the diarization performance remains consistent on speech recorded with distant or close-talking microphones.

Further improvements can include the use of a re-segmentation procedure, the adoption of VBx Bayesian HMM, the exploration of iterative clustering approaches with embeddings from longer segments in a second pass, and  re-visiting automated ways to estimate the number of speakers in a recording such as NME-SC.

\section{Acknowledgements}
We would like to thank Yoshua Bengio, Samuele Cornell, and the rest of the SpeechBrain developers for the helpful suggestions.

\bibliographystyle{IEEEtran}

\bibliography{references}

\end{document}